\documentclass[aps,reprint,twocolumn,superscriptaddress]{revtex4-1}
\usepackage[colorlinks,bookmarks=true,citecolor=blue,linkcolor=red,urlcolor=blue]{hyperref}
\usepackage{amsmath,amsfonts,amssymb,mathtools}
\usepackage{mathrsfs}
\usepackage{graphicx,float}
\usepackage{multirow}
\usepackage{makecell}
\usepackage[ruled,vlined]{algorithm2e}
\usepackage{algorithmic}
\usepackage{color}
\usepackage{dcolumn}
\usepackage{bm}
\usepackage{subfiles}
\usepackage{verbatim} 
\usepackage{tikz}
\usepackage{longtable}
\usepackage{rotating} 
\usepackage{float}

\usepackage[T1]{fontenc}
\usepackage[utf8]{inputenc}
\usepackage{lmodern} 

\usepackage[title]{appendix}
\usepackage{url}
\usepackage{booktabs}

\begin{document}

\title{Magnetic bulk photovoltaic effect as a probe of magnetic structures of $\rm{EuSn_2As_2}$}

\author{Hanqi Pi}
\affiliation{Beijing National Laboratory for Condensed Matter Physics and Institute of physics, Chinese academy of sciences, Beijing 100190, China}
\affiliation{University of Chinese academy of sciences, Beijing 100049, China}

\author{Shuai Zhang}
\affiliation{Beijing National Laboratory for Condensed Matter Physics and Institute of physics, Chinese academy of sciences, Beijing 100190, China}
\affiliation{University of Chinese academy of sciences, Beijing 100049, China}

\author{Hongming Weng}
\email{hmweng@iphy.ac.cn} 
\affiliation{Beijing National Laboratory for Condensed Matter Physics and Institute of physics, Chinese academy of sciences, Beijing 100190, China}
\affiliation{University of Chinese academy of sciences, Beijing 100049, China}
\affiliation{Songshan Lake Materials Laboratory, Dongguan, Guangdong 523808, China}

\date{\today}
	
\begin{abstract}
The bulk photovoltaic effect (BPVE) is a second-order optical process in noncentrosymmetric materials that converts the light into DC currents. BPVE is classified into shift current and injection current according to the generation mechanisms, whose dependence on the polarization of light is sensitive to the spatial and time-reversal symmetry of materials. In this work, we present a comprehensive study on the BPVE response of $\mathrm{EuSn_2As_2}$ with different magnetic structures through symmetry analysis and first-principles calculation. We demonstrate that the interlayer antiferromagnetic (AFM) $\mathrm{EuSn_2As_2}$ of even-layer breaks the inversion symmetry and has the second-order optical responses. Moreover, the bilayer AFM $\mathrm{EuSn_2As_2}$ not only displays distinct BPVE responses when magnetic moments align in different directions, but also shows symmetry-related responses in two phases which have mutually perpendicular in-plane magnetic moments. Due to the dependence of BPVE responses on the polarization of light and magnetic symmetry, these magnetic structures can be distinguished by the circular polarized light with well-designed experiments. Our work demonstrates the feasibility of the BPVE response as a tool to probe the magnetic structure.
\end{abstract}
	
\maketitle
	
\section{\label{sec:level1}Introduction}
Nonlinear optical phenomena play an essential role in condensed matter physics to advance fundamental knowledge about materials and stimulate the development of technological applications. For instance, the second harmonic generation (SHG) has been applied to probe the electronic, magnetic and crystallographic structures of materials\cite{nvemec2018antiferromagnetic,sun2019giant,fiebig2005second, li2013probing,heide2022probing,vampa2015all,luu2018measurement,denev2011probing}. The high-harmonic generation (HHG) recently has been intensively studied to reveal the electron dynamics\cite{silva2018high}, band structures\cite{lv2021high} and topological phase transitions\cite{qian2022role}. Among various nonlinear optical processes, the bulk photovoltaic effect (BPVE) in noncentrosymmetric materials converting light into a DC current has gained numerous interests\cite{belinicher1980photogalvanic,von1981theory,fridkin2001bulk,moore2010confinement,nastos2006optical,xu2020comprehensive,ni2021giant}. As it surpasses the Shockley-Queisser limit and generates above-band-gap photovoltages,  the BPVE is expected to be a replacement for the conventional solar cell\cite{tan2016shift,cook2017design}.
Besides, BPVE can be utilized to develop promising photodetectors because it does not require a bias voltage which would cause the dark current.
Furthermore, due to the close relationship with geometric quantities such as Berry connection and Berry curvature\cite{morimoto2016topological,nagaosa2017concept,orenstein2021topology}, the BPVE is employed to obtain the band geometry and 
topology information of materials\cite{de2017quantized,ma2017direct,chan2017photocurrents}.
 
 The photocurrent in BPVE consists of the shift current and the injection current, which are induced by the change in the charge center and group velocity of electrons during the interband transitions, respectively\cite{sipe2000second,nagaosa2017concept}. The study on the BPVE photocurrents has been confined to nonmagnetic materials for decades\cite{young2012first,tan2016enhancement,wu2017giant,sipe2000second,xu2020comprehensive,sturman2021photovoltaic}. In systems with time-reversal symmetry ($\mathcal{T}$), these two currents 
could be distinguished by their dependence on the polarization of incident photons. While the injection current is excited only by the circular polarized light, the shift current can be generated by light regardless of polarization\cite{sipe2000second,ahn2020low}. Besides, the circular injection current alters its direction when the helicity of incident light changes while the linear shift current does not\cite{wang2019ferroicity}. 
 
 In recent years, a series of works\cite{zhang2019switchable,wang2020electrically,holder2020consequences,ahn2020low,watanabe2021chiral} have shown the growing interests in the BPVE response of magnetic materials, which has a contrast dependence on the polarization of light compared to nonmagnetic ones. In magnetic materials where $\mathcal{T}$ is broken and the space-time inversion symmetry $\mathcal{PT}$ is preserved, shift current excited by the linear polarized light is prohibited while injection current excited by light of any polarization is allowed. It results in the linear injection current and circular shift current in the $\mathcal{PT}$ magnetic materials\cite{zhang2019switchable,wang2020electrically}. Moreover, in system with neither $\mathcal{T}$ nor $\mathcal{PT}$, both BPVE photocurrents can be induced by either linear or circular polarized light\cite{ahn2020low}.
 
 Due to the sensitivity to the crystalline symmetry and the polarization of light, BPVE responses could be employed to probe the magnetic structure of materials. We take the magnetic material $\mathrm{EuSn_2As_2}$ as an example and analyze the BPVE in three magnetic structures of $\mathrm{EuSn_2As_2}$. The symmetry analysis demonstrates that BPVE is allowed to exist only in even-layer $\mathrm{EuSn_2As_2}$ with antiferromagnetic (AFM) order. BPVE in three AFM phases of bilayer $\mathrm{EuSn_2As_2}$ are investigated through group representation theory and first-principles calculation. The result shows that both linear injection current and circular shift current can be generated in AFM bilayer $\mathrm{EuSn_2As_2}$ with the in-plane magnetic moments, while only linear injection current exist when the magnetic moments are out of plane. Besides, the two magnetic structures with mutually perpendicular in-plane magnetic moments has the similar BPVE responses, which can be explained by the connection between their magnetic symmetry groups, namely their magnetic moments alignment is related by the $C_{4z}$ rotation. By exploiting the discrepancies among their nonlinear optical responses, we provide an experimental protocol to distinguish the magnetic structures by two circular polarized light beams with the opposite helicity.
	
	\section{Results}
	\subsection{The bulk photovoltaic effect}
	When illuminated by the monochromatic light $\boldsymbol{\tilde{E}}(t)=\boldsymbol{E}(\omega)e^{-i \omega t}+\operatorname{c.c}$, noncentrosymmetric materials can give rise to a DC photocurrent described by the following relation\cite{sipe2000second},
		\begin{equation}\label{definition}
		\begin{split}
			J^a=&\sigma^{abc}(0;-\omega,\omega) E_b(-\omega)E_c(\omega)\\
			+&\sigma^{abc}(0;\omega,-\omega) E_b(\omega)E_c(-\omega)\\
                =&\sigma^{abc}(0;-\omega,\omega) E_b(-\omega)E_c(\omega)+\operatorname{c.c.}\\
			=&2\sigma^{abc}(0;-\omega,\omega) E_b(-\omega)E_c(\omega),
		\end{split}
	\end{equation}
	where we use the intrinsic permutation symmetry of nonlinear optical coefficients to derive the final expression. Different from the photovoltaic effect in a $p$-$n$ junction, the direct current generated in noncentrosymmetric homogeneous crystals comes from the second-order optical process and is termed as bulk photovoltaic effect (BPVE) or photogalvanic effect (PGE)\cite{sturman2021photovoltaic}.  
	
	As the photocurrent is a real quantity and $E_b(-\omega)=E_b^*(\omega)$, we take the complex conjugate of (\ref{definition}) and obtain $\left[\sigma^{abc}(0;-\omega,\omega)\right]^*=\sigma^{acb}(0;-\omega,\omega)$. Therefore, the real/imaginary part of $\sigma^{abc}(0;-\omega,\omega)$ is symmetric/antisymmetric to the exchange of the latter two index,
	\begin{equation}
		\begin{aligned}
			&\operatorname{Re}\sigma^{abc}(0;-\omega,\omega)=\operatorname{Re}\sigma^{acb}(0;-\omega,\omega)\\
			&\operatorname{Im}\sigma^{abc}(0;-\omega,\omega)=-\operatorname{Im}\sigma^{acb}(0;-\omega,\omega),
		\end{aligned}
	\end{equation}
	We denote $\operatorname{Re}\sigma(0;-\omega,\omega)$  as $\sigma_L$ and $i\operatorname{Im}\sigma(0;-\omega,\omega)$  as $i\sigma_C$.  With this notation, the definition of BPVE (\ref{definition}) can be reformulated as\cite{de2017quantized,rees2021direct},
		\begin{equation}\label{modified version of BPVE def}
		\begin{aligned}
			J^a=	2&\sigma_L^{abc}E^*_bE_c+2i\sigma_C^{ad}[\boldsymbol{E^*\times E}]_d.
		\end{aligned}
	\end{equation}
    Because $\boldsymbol{E^*\times E}$ is required to be imaginary, the photocurrent described by $\sigma_C$ can only be contributed by the circular polarized light and is called the circular photogalvanic effect (CPGE). On the contrary, PGE described by $\sigma_L$ is called the linear photogalvanic effect (LPGE). Nonetheless, LPGE can be generated under the illumination of either linear or cicular polarized light. Moreover, because the change of helicity of circular polarized light is equivalent to the exchange of $\boldsymbol{E^*}$ and $\boldsymbol{ E}$, the photocurrent of CPGE will alter its direction when the circular polarized light reverses its helicity while the LPGE does not\cite{sturman2021photovoltaic}.

	Within the framework of perturbation theory and only considering the interband transitions, we can obtain the expression of BPVE described by the following two parts\cite{sipe1993nonlinear,aversa1995nonlinear,sipe2000second},
	 \begin{equation}\label{injection current}
		\begin{aligned}
			\frac{dJ^a_{\text{inject}}}{dt}&=2\eta_{\text{inject}}^{a b c}\left(0 ; -\omega,\omega\right) E_{b}\left(-\omega\right) E_{c}\left(\omega\right)
			\\&=-\frac{2e^{3}\pi}{ \hbar^{2}}\sum_{n m \boldsymbol{k}}\Delta_{mn}^{a} r_{nm}^{b}r_{mn}^{c} f_{ n m } \delta\left(\omega_{mn}-\omega\right)\\
			&\times E_{b}\left(-\omega\right) E_{c}\left(\omega\right)
				\end{aligned}
		\end{equation}
		 \begin{equation}\label{shift current}
		\begin{aligned}
			J^a_{\text{shift}}&=2\sigma_{\text{shift}}^{a b c}\left(0 ; -\omega,\omega\right) E_{b}\left(-\omega\right)E_{c}\left(\omega\right)
			\\&=-\frac{ie^{3}\pi}{\hbar^{2}}\sum_{nm\boldsymbol{k}} f_{n m} \left[ \left(r_{mn}^{c}\right)_{;k ^a} r_{nm}^{b}-\left(r_{n m}^{b}\right)_{;k ^a} r_{m n}^{c}\right]
			\\ &
			\times\delta(\omega_{mn}-\omega)E_{b}\left(-\omega\right)E_{c}\left(\omega\right)
			\\&=-\frac{e^{3}\pi}{\hbar^{2}}\sum_{nm\boldsymbol{k}} f_{n m}r _ { n m } ^ { b } r _ { m n } ^ { c } \left( \mathcal{R}_{mn;a}^c-\mathcal{R}_{nm;a}^b\right)\\
			&\times\delta(\omega_{mn}-\omega)E_{b}\left(-\omega\right)E_{c}\left(\omega\right),
		\end{aligned}
	\end{equation}
	where $\Delta_{mn}^{a}\equiv v_m^a-v_n^a$ and $\hbar\omega_{mn}\equiv E_m-E_n$ denotes the group velocity along $a$ and eigen-energy difference between band $m$ and $n$, respectively. $r_{nm}^b=\left\langle u_{n\boldsymbol{k}}\right| i\partial_{k^b}\left| u_{n\boldsymbol{k}}\right\rangle$ is the transition dipole moment along $b$ between band $m$ and $n$ where $\left| u_{n\boldsymbol{k}}\right\rangle$ is the periodic part of Bloch wavefunction\cite{blount1962formalisms}. $f_n$ is the Fermi-Dirac distribution of band $n$ and $f_{nm}\equiv f_n-f_m$. $ \left(r_{mn}^{c}\right)_{;k ^a} \equiv{\partial_a r^c_{mn}}-i r^c_{mn}\left({A}^a_{mm}-A^a_{nn}\right)$ is the generalized derivative and $A_{nn}^a=\left\langle u_{n\boldsymbol{k}}\right| i\partial_{k^a}\left| u_{n\boldsymbol{k}}\right\rangle$ is the Abelian Berry connection. $\mathcal{R}_{mn;a}^c\equiv i\frac{\partial \ln r_{mn}^c}{\partial_{k^a}}+A^a_{mm}-A^a_{nn}$ is the shift vector indicating the change of electron position in real space when transiting from band $n$ to $m$. Note that we adopt the definition of shift vector in Ref.\cite{ahn2020low} which is different the one in nonmagnetic systems\cite{sipe2000second,tan2016enhancement,wang2020electrically}.
	
	We could see that (\ref{injection current}) describes a DC current generated from the change of electron group velocity that grows linearly with time. In real materials, however, the photocurrent will saturate after a time $\tau$ because of the impurities, defects, and other scattering mechanisms. Therefore, the photocurrent in (\ref{injection current}) is named injection current. Similarly, the DC current in (\ref{shift current}) originating from the change of electron position in real space is called shift current. According to their dependence on the polarization of light as mentioned above, the two photocurrents can be further classified into linear/circular injection current and linear/circular shift current. We obtain the corresponding response coefficients by separating the real and imaginary part of $\eta_{\text{inject}}$ and $\sigma_{\text{shift}}$,
	\begin{equation}\label{coefficients of injection current}
	\begin{aligned}
		2\eta^{abc}_{\text {LI}}(0;-\omega,\omega)=&
		-\frac{e^{3}\pi}{ \hbar^{2}}\sum_{n m \boldsymbol{k}}f _ { n m }\Delta _ { mn } ^ { a }\left\{r_{nm}^{b},r_{mn}^{c}\right\} \\ &\times\delta\left(\omega_{mn}-\omega\right)\\
		2\eta^{abc}_{\text {CI }}(0;-\omega,\omega)=&
		\frac{ie^{3}\pi}{ \hbar^{2}}\sum_{n m \boldsymbol{k}}f _ { n m }\Delta _ { mn } ^ { a }\left[r_{nm}^{b},r_{mn}^{c}\right]\\ &\times \delta\left(\omega_{mn}-\omega\right)
			\end{aligned}
	\end{equation}
	\begin{equation}\label{coefficients of shift current}
	\begin{aligned}
		2\sigma^{abc}_{\text {LS}}(0;-\omega,\omega)=&
		-\frac{ie^{3}\pi}{2\hbar^{2}}\sum_{nm\boldsymbol{k}} f_{n m}	\left\{\left(r_{mn}^{c}\right)_{;k ^a}, r_{nm}^{b}\right\}\\
		&\times\left[\delta\left(\omega_{mn}-\omega\right)-\delta\left(\omega_{mn}+\omega\right)\right]\\
		2\sigma^{abc}_{\text {CS}}(0;-\omega,\omega)=&	-\frac{e^{3}\pi}{2\hbar^{2}}\sum_{nm\boldsymbol{k}} f_{n m}\left[\left(r_{mn}^{c}\right)_{;k ^a}, r_{nm}^{b}\right]\\
		&\times\left[\delta\left(\omega_{mn}-\omega\right)-\delta\left(\omega_{mn}+\omega\right)\right],
	\end{aligned}
\end{equation}
	 where we have used the relations 
	\begin{equation}
			\left(r_{nm}^{b}\right)^*=r_{mn}^{b},\quad \left[ \left(r_{mn}^{b}\right)_{;k ^a}\right]^*=\left(r_{nm}^{b}\right)_{;k ^a},
	\end{equation}
	and defined the commutators and anticommutators,
	\begin{equation}
		\begin{aligned}
		\left\{r_{nm}^{b},r_{mn}^{c}\right\}&\equiv\left(r_{nm}^{b}r_{mn}^{c}+r_{nm}^{c}r_{mn}^{b}\right)\\
		\left[r_{nm}^{b},r_{mn}^{c}\right]&\equiv\left(r_{nm}^{b}r_{mn}^{c}-r_{nm}^{c}r_{mn}^{b}\right)\\
			\left\{\left(r_{mn}^{c}\right)_{;k ^a}, r_{nm}^{b}\right\}&\equiv\left[	\left(r_{mn}^{c}\right)_{;k ^a} r_{nm}^{b}+\left(r_{mn}^{b}\right)_{;k ^a} r_{nm}^{c}\right]\\
			\left[\left(r_{mn}^{c}\right)_{;k ^a}, r_{nm}^{b}\right]&\equiv\left[	\left(r_{mn}^{c}\right)_{;k ^a} r_{nm}^{b}-\left(r_{mn}^{b}\right)_{;k ^a} r_{nm}^{c}\right].
		\end{aligned}
	\end{equation}
 They are the same with the expressions in Ref.\cite{wang2020electrically}.
	
	Under the spatial symmetry operations, response coefficients in (\ref{coefficients of injection current}) and (\ref{coefficients of shift current}) obey the same transformation rule as a third-rank tensor and thus vanish in systems with the spatial inversion symmetry $\mathcal{P}$. As the Neumann's principle breaks down in the dynamic processes such as the transport phenomenon\cite{birss1963macroscopic}, the transformation rule of response tensors under $\mathcal{T}$ can not be acquired by applying $\mathcal{T}$ to the relevant physical quantities, i.e., electric current $\boldsymbol{J}$ and electric field $\boldsymbol{E}$. However, we can obtain the transformation rule of $\eta_{\text{inject}}$ and $\sigma_{\text{shift}}$ from their expressions (\ref{coefficients of injection current},\ref{coefficients of shift current}). In systems with time-reversal symmetry,  $\boldsymbol{r}_{mn}(\boldsymbol{k}) =\boldsymbol{r}_{nm}(-\boldsymbol{k})$ and $\left(\boldsymbol{r}_{mn}\right)_{;\boldsymbol{k}}(\boldsymbol{k}) =\left(\boldsymbol{r}_{nm}\right)_{;\boldsymbol{k}}(\boldsymbol{-k}) $. Therefore, the circular shift and linear injection current vanish in nonmagnetic system with $\mathcal{T}$, while the linear shift and circular injection current are allowed to exist. In the noncentrosymmetric magnetic systems with preserved combined $\mathcal{PT}$ symmetry, the linear shift and circular injection current vanish while the other two photocurrents, namely circular shift and linear injection current, exist\cite{ahn2020low}. In the next section, we take $\mathrm{EuSn_2As_2}$ as an example to analyze the BPVE responses in magnetic structures with symmetry analysis and first-principles calculation.
	
	\subsection{Crystal structure and symmetry analysis of  bilayer $\mathrm{EuSn_2As_2}$}
	The bulk $\mathrm{EuSn_2As_2}$ has a layered structure and belongs to space group $R\overline{3}m$. Every layer component is composed of a trigonal Eu layer sandwiched by two SnAs layers. $\mathrm{EuSn_2As_2}$ was found to be an axion insulator in the A-type AFM phase belonging to magnetic space group (MSG) $R\overline{3}m'$ and a strong topological insulator in the paramagnetic phase \cite{li2019dirac}. BPVE is prohibited in the two phases because they both have $\mathcal{P}$  with the inversion center on Eu atoms. When we confine the $\mathrm{EuSn_2As_2}$ to have a collinear magnetic structure, only the even-layer $\mathrm{EuSn_2As_2}$ in the A-type AFM phase breaks $\mathcal{P}$ and is able to generate BPVE photocurrents.
	\begin{figure}[htbp] 
		\centering 
			\includegraphics[width=0.5\textwidth]{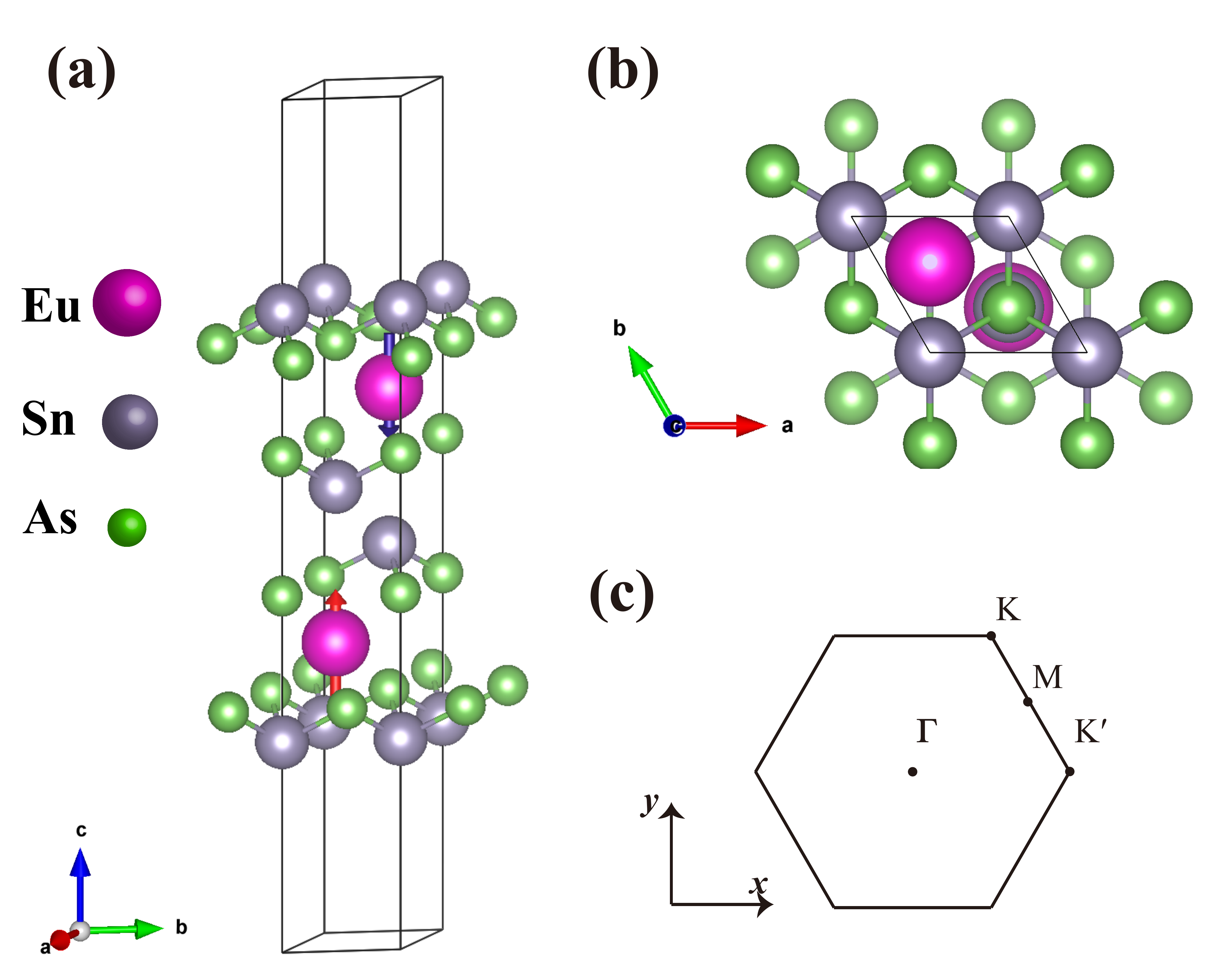}
		\caption{\label{fig1} (a-b) The crystal and magnetic stucture of bilayer AFM-$z$ $\mathrm{EuSn_2As_2}$. The crystal axes $a$ is along the $x$ direction in the Cartesian
        coordinate system. (c) The Brillouin zone of bilayer $\mathrm{EuSn_2As_2}$.}
	\end{figure}

 \begin{figure*}[htb]
		\centering 
		\includegraphics[width=0.8\textwidth]{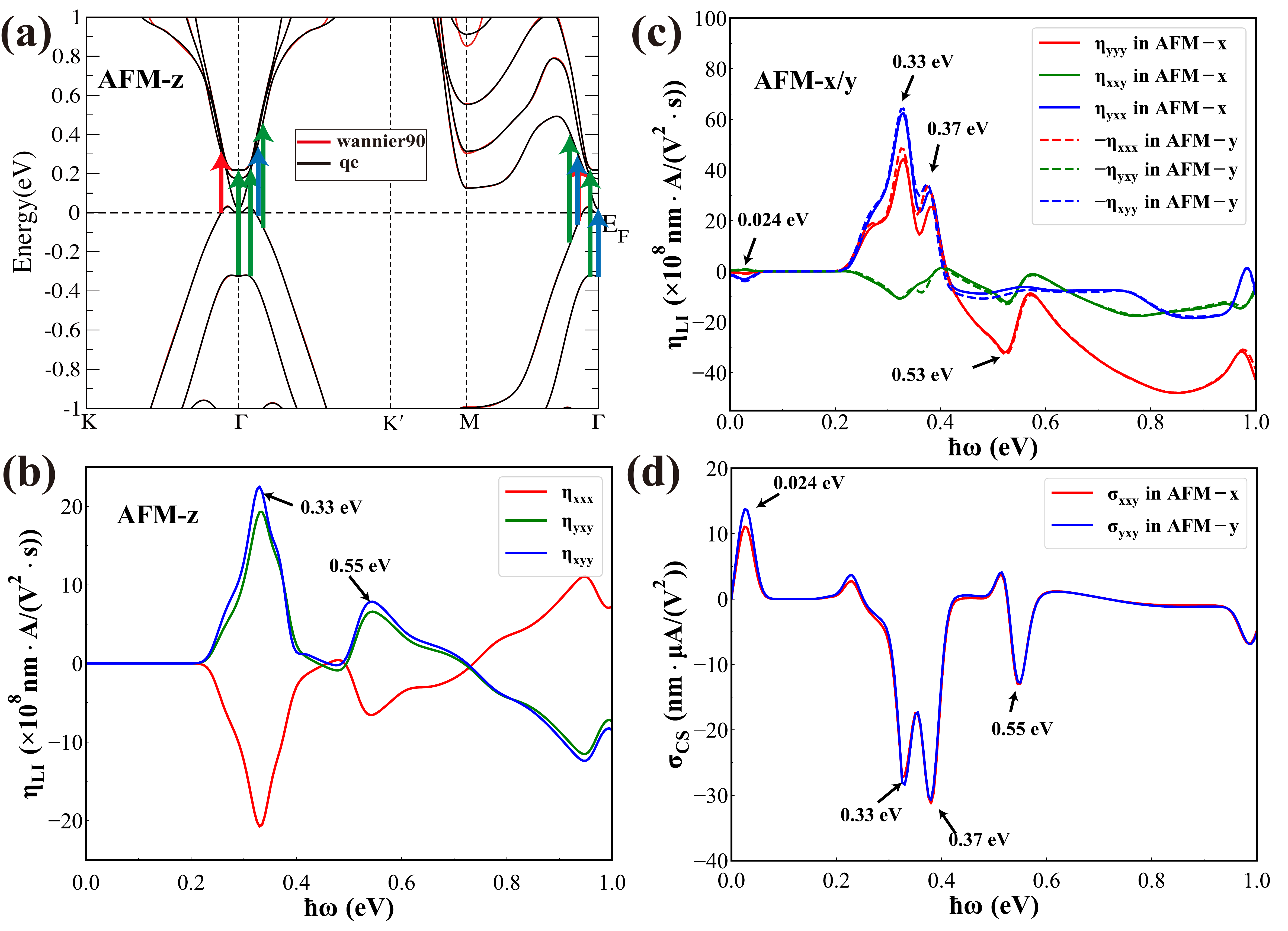}
		\caption{(a) The band structures of bilayer AFM-$z$ $\mathrm{EuSn_2As_2}$ with spin-orbit coupling (SOC) calculated with first-principles calculation (black) and tight-binding Hamiltonian based on the Wannier functions (red). The red, blue and green arrows marked the interband contributions to peaks in (b-c) centering at 0.33 eV, 0.37 eV and 0.55 eV, respectively. (b) The linear injection responses $\eta_{\text{LI}}$ in AFM-$z$ phase. (c) The linear injection responses $\eta_{\text{LI}}$ in AFM-$x/y$ phases. (d) The circular shift responses $\sigma_{\text{CS}}$ in AFM-$x/y$ phases. \label{fig2}}
	\end{figure*}

	We consider three AFM structures of the bilayer $\mathrm{EuSn_2As_2}$ with the magnetic moments along the $\hat{x}$ (AFM-$x$),  $\hat{y}$ (AFM-$y$), and $\hat{z}$ (AFM-$z$) directions.  The nonvanishing independent components of the BPVE response tensors in those magnetic structures can be obtained by utilizing their properties under $\mathcal{T}$ and the irreducible representations (IRREPs) of the crystalline point group (CPG). For instance, the crystal structure of bilayer AFM-$z$ $\mathrm{EuSn_2As_2}$ is illustrated in Fig. \ref{fig1}(a), where the magnetic moments break $\mathcal{P}$ but preserve $\mathcal{PT}$. It belongs to the magnetic point group (MPG) $\overline{3}'m'$, $\mathcal{M}=\left\{ \mathcal{I},2C_3, 3C_2,\mathcal{PT},2S_6\mathcal{T},3\sigma_d\mathcal{T}\right\}$, which can be decomposed into two forms, $\mathcal{M}=D_3\oplus\left(D_{3d}-D_3\right)\mathcal{T}$ and $\mathcal{M}=D_3\otimes\{\mathcal{I},\mathcal{PT}\}$. As $\sigma_{\text{CI}}$ and $\eta_{\text{LS}}$ are invariant under $\mathcal{T}$, the effect of $\mathcal{T}$ is equivalent to the identity operation and we can acquire their nonvanishing independent components by analyzing the IRREPs and generators of CPG $D_{3d}$. The $\mathcal{P}$ in $D_{3d}$ prohibits the existence of $\sigma_{\text{CI}}$ and $\eta_{\text{LS}}$, being consistent with the former conclusion that the linear shift and circular injection current vanish in systems with $\mathcal{PT}$. Similarly, we can analyze $\sigma_{\text{CS}}$ and $\eta_{\text{LI}}$ with IRREPs of CPG $D_{3}$ due to their invariance under $\mathcal{PT}$. As the bilayer $\mathrm{EuSn_2As_2}$ is a quasi-2D material, we focus on the in-plane responses. Take the electric current $\boldsymbol{J}$ as an example, we only concern the $J_x$ and $J_y$ components that transform as $E$ in $D_3$ as illustrated in Table \ref{characters of D3}. Table \ref{characters of D3} presents the IRREPs of group $D_3$ and characters of the second-rank tensor for the group elements\cite{lax2001symmetry}. As $\sigma_{\text{CS}}$ ($\eta_{\text{LI}}$) relates a polar vector $\boldsymbol{J}$ and an asymmetric (symmetric) second-order tensor $[\boldsymbol{EE^*}]_{\text{asym}}$ ($[\boldsymbol{EE^*}]_{\text{sym}}$), we obtain the representation of $\sigma_{\text{CS}}$ ($\eta_{\text{LI}}$) and decompose it into the direct sum of irreducible representations as followings,
	
		\begin{equation}
		\begin{split}
			&\Gamma_{\sigma_{\text{CS}}}=\Gamma_{J}\otimes\Gamma_{(EE^{*})_{\text{asym}}}=E,\\
			&\Gamma_{\eta_{\text{LI}}}=\Gamma_{J}\otimes\Gamma_{(EE^{*})_{\text{sym}}}=A_1\oplus A_2\oplus 2E.
		\end{split}
	\end{equation}
	Because $\Gamma_{\eta_{\text{LI}}}$ includes one identity representation, there is one nonvanishing independent component in ${\eta_{\text{LI}}}$. Utilizing the symmetry confinement of generators, $C_{2x}$ and  $C_{3z}$, we obtain the nonvanishing independent components, i.e., $-\eta_{\text{LI}}^{xxx}=\eta_{\text{LI}}^{xyy}=\eta_{\text{LI}}^{yxy}$. On the contrary, the decomposition of $\Gamma_{\sigma_{\text{CS}}}$ without the identity representation indicates that the circular shift response vanishes in the bilayer AFM-$z$ $\mathrm{EuSn_2As_2}$.
	\begin{table}[H]
		\centering
		\begin{tabular}{c|ccc|c}
			\toprule
			$D_3$  & $E$ & $2C_3$ & $3C_2$& basis function  \\
			\hline
			$A_1$ & 1 & 1 & 1&\\
			$A_2$ & 1 & 1 & -1&$z,\ R_z$\\
			$E$ & 2 & -1 & 0&$(x,y),\ (R_x,R_y) $\\
			\hline
			$\Gamma_{(EE^{\star})_{\text{sym}}}$ & 3 & 0 & 1&\\
			$\Gamma_{(EE^{\star})_{\text{asym}}}$ & 1 & 1 & -1&\\
			\bottomrule
		\end{tabular}
		\caption{\label{characters of D3}Characters for irreducible representations of group $D_3$ and characters of the symmetric and asymmetric second-rank tensor for group elements. }
	\end{table}

        \begin{figure*}[ht]
		\centering 
    	\includegraphics[width=1\textwidth]{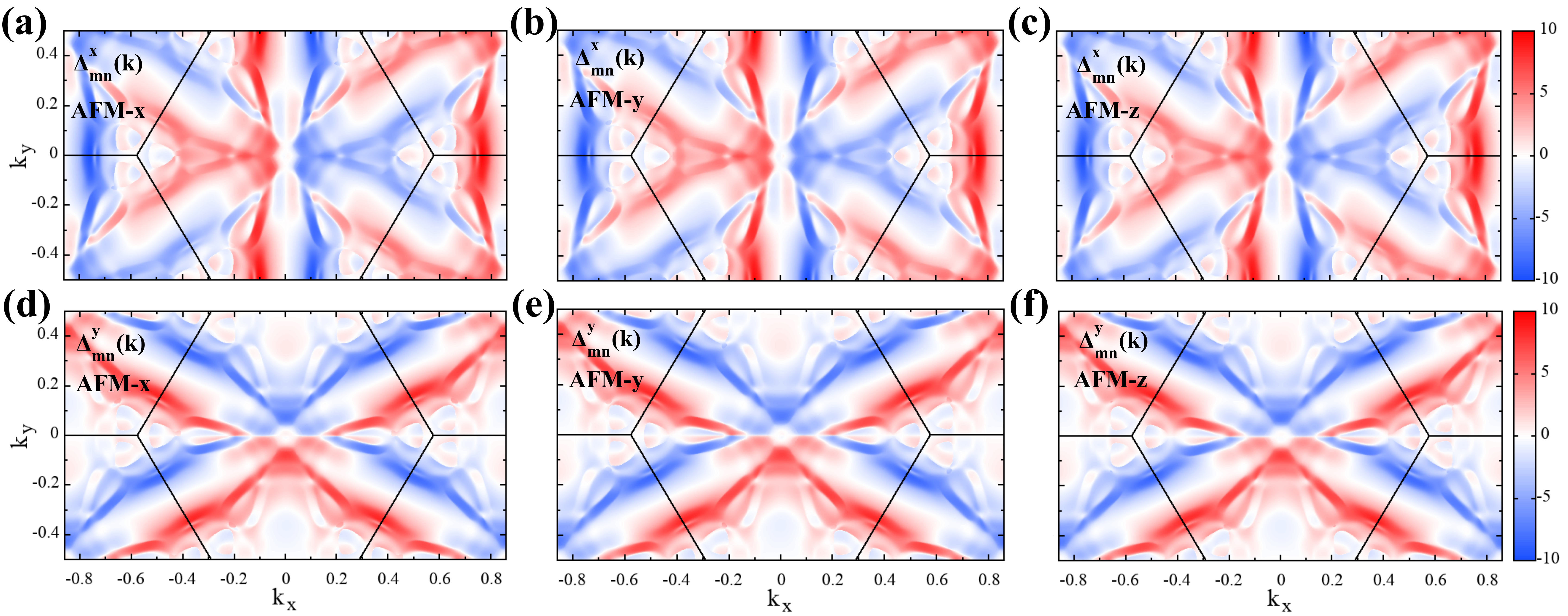}
		\caption{ The distribution of the group velocity difference $\Delta_{m n}^x(\mathbf{k})$ and $\Delta_{m n}^y(\mathbf{k})$ in the BZ of (a,d) AFM-$x$ , (b,e) AFM-$y$ and (c,f) AFM-$z$ structures .\label{fig3}}
	\end{figure*}
		
	We can analyze the BPVE response in bilayer AFM-$x$ and AFM-$y$ $\mathrm{EuSn_2As_2}$ through the similar scheme. The bilayer AFM-$x$ $\mathrm{EuSn_2As_2}$ belongs to MPG $2'/m$ that can be decomposed into $\mathcal{M}=C_s\oplus\left(C_{2h}-C_s\right)\mathcal{T}$ and $\mathcal{M}=C_s\otimes\{\mathcal{I},\mathcal{PT}\}$, while the AFM-$y$ phase belongs to the MPG $2/m'$ that can be decomposed into $\mathcal{M}=C_2\oplus\left(C_{2h}-C_2\right)\mathcal{T}$ and $\mathcal{M}=C_2\otimes\{\mathcal{I},\mathcal{PT}\}$. The character tables for the mentioned CPGs and detailed analysis can be found in Supplementary Material. In AFM-$x$ and AFM-$y$ phases, $\sigma_{\text{CI}}$ and $\eta_{\text{LS}}$ are still prohibited. However, the lowered symmetry resulted from the in-plane magnetic moments lead to more BPVE responses. To be more specific, the breakdown of $C_{3z}$ symmetry not only gives rise to the violation of equivalence relations among in-plane components, i.e., $-xxx=xyy=yxy$, $-yyy=xxy=yxx$, but also allows for the existence of in-plane circular shift current, i.e., ${\sigma^{xxy}_{\text{CS}}}$ in AFM-$x$ and ${\sigma^{yxy}_{\text{CS}}}$ in AFM-$y$.

	The distinct BPVE response in different MSG helps us detect the magnetic structure of $\mathrm{EuSn_2As_2}$. First of all, as the nonvanishing second-order optical responses only exists in even-layer AFM $\mathrm{EuSn_2As_2}$, we could easily distinguish it from odd number layer structures. Secondly, because the circular shift current only exist in AFM-$x$ and AFM-$y$ phases and it reverses direction when the incident circular polarized light changes helicity, we can measure the difference between response currents of circular polarized light with opposite helicity, $\vec{J}_{\circlearrowleft}-\vec{J}_{\circlearrowright}$, to distinguish them from the AFM-$z$ phase. Finally, we could differentiate the AFM-$x$ and AFM-$y$ phase by a beam of light regardless of the polarization. When illuminated by linear polarized light with polarization along $x$ or $y$ direction, the response photocurrent flows perpendicular to the magnetic moments due to the nonzero $\eta_{\mathrm {LI }}^{yxx,yyy}$ and $\eta_{\mathrm {LI }}^{xxx,xyy}$ in the AFM-$x$ and AFM-$y$ phase, respectively. When illuminated by circular polarized light propagating in $\hat{z}$ direction, the response photocurrent flows parallel to the magnetic moments due to the nonzero $\eta_{\mathrm {LI }}^{xxy}, {\sigma^{xxy}_{\text{CS}}}$ and $\eta_{\mathrm {LI }}^{yxy}, {\sigma^{yxy}_{\text{CS}}}$ in the AFM-$x$ and AFM-$y$ phase, respectively.
	\subsection{First-principles calculation of BPVE in bilayer $\mathrm{EuSn_2As_2}$}
	
	We obtain the band structures of the bilayer AFM-$x/y/z$ $\mathrm{EuSn_2As_2}$ by first-principles calculation as shown in Fig.\ref{fig2}(a) and Fig.\ref{figs1}(a,b). They have similar electronic structures with a narrow bandgap of 15 meV, indicating that the direction of magnetic moments has negligibly small influence on the band structure but has direct constrain on their BPVE responses. The linear injecton responses of the three magnetic phases are presented in Fig.\ref{fig2}(b-c). As discussed before, there are three nonvanishing components of $\eta_{\mathrm {LI }}$ in AFM-$z$ phase, which are supposed to share the same magnitude and only differ by a sign. The little inconsistancy shown in Fig.\ref{fig2}(b) is attributed to the violation of symmetry in generating Wannier functions. On the contrary,  the three components in AFM-$x/y$ phases are independent as demonstrated in Fig.\ref{fig2}(c) due to the absence of $C_{3z}$. Furthermore, $\mathrm{EuSn_2As_2}$ in AFM-$x/y$ phases display stronger linear injection response. Fig.\ref{fig2}(d) shows the circular shift response in AFM-$x/y$ phase, which is absent in AFM-$z$ phase because of the relations ($-xxx=xyy=yxy, -yyy=yxx=xxy$) required by $C_{3z}$ and the antisymmetric property of circular shift conductivity to the last two indices.
        \begin{figure*}[ht]
		\centering 
    	\includegraphics[width=1\textwidth]{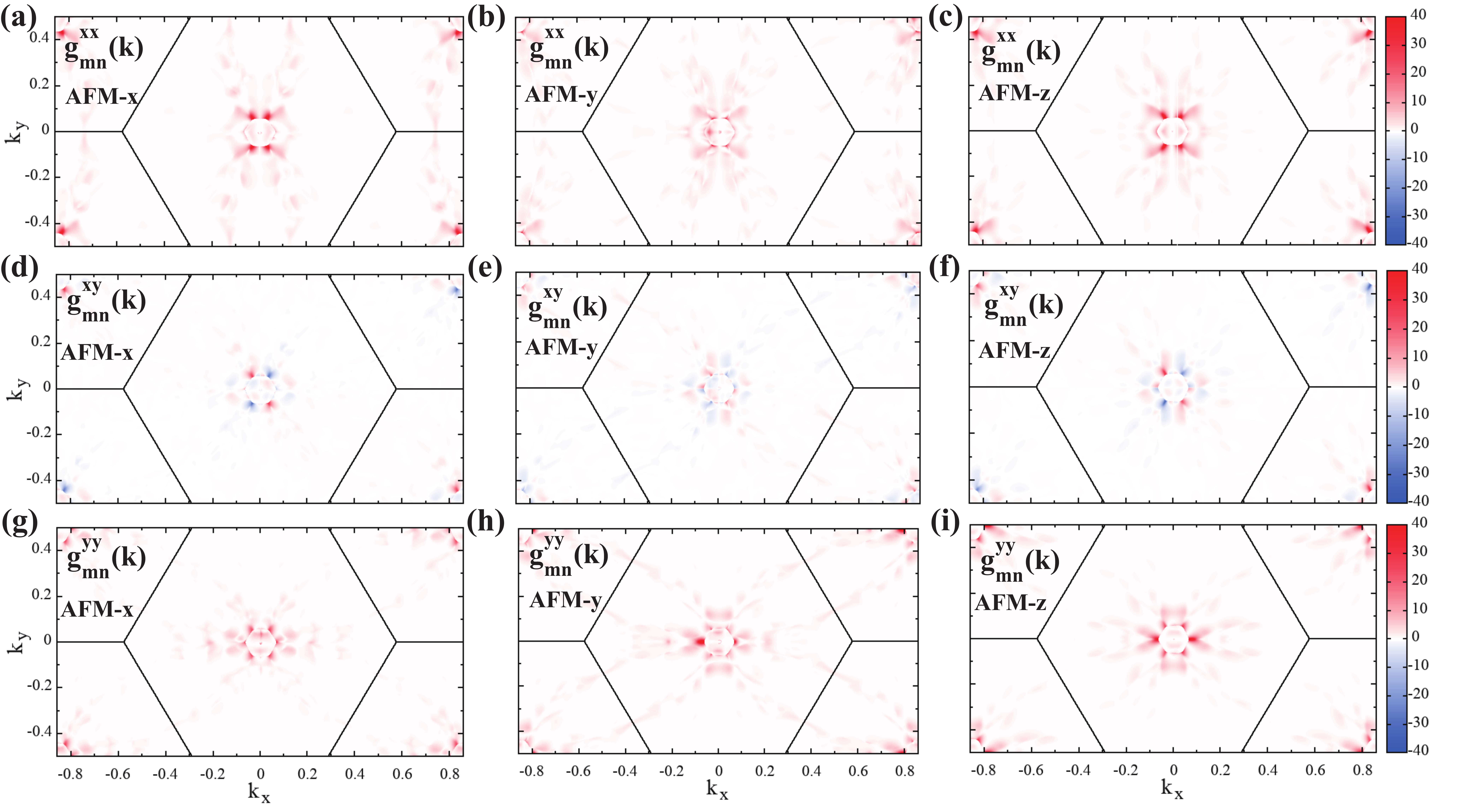}
		\caption{ The distribution of quantum metric $g_{mn}^{bc}(\mathbf{k})$ in the BZ of (a,d,g) AFM-$x$, (b,e,h) AFM-$y$, and (c,f,i) AFM-$z$ structures.\label{fig4}}
	\end{figure*}
 
	 From Fig.\ref{fig2}(c-d), we notice that there is a connection between BPVE responses in AFM-$x$ and AFM-$y$ phase, i.e., $\eta_{yyy}^{\text{AFM-x}}=-\eta_{xxx}^{\text{AFM-y}}$, $\eta_{xxy}^{\text{AFM-x}}=-\eta_{yxy}^{\text{AFM-y}}$ , $\eta_{yxx}^{\text{AFM-x}}=-\eta_{xyy}^{\text{AFM-y}}$ and $\sigma_{xxy}^{\text{AFM-x}}=\sigma_{yxy}^{\text{AFM-y}}$, with minor differences resulted from the little violated symmetry in generating the Wannier functions and the tiny magnetic anisotropy. We attribute it to be a consequence required by symmetry rather than a coincidence. As the bilayer $\mathrm{EuSn_2As_2}$ is a quasi-2D material, the effect of $M_x$ in MSG $P2'/m$ is identical to $C_{2y}$ when constranied in the $x-y$ plane. Therefore, the MSG $P2/m'$ of AFM-$y$ phase can be transformed into the MSG $P2'/m$ of AFM-$x$ phase when transformed under $C_{4z}$. Applied with $C_{4z}$, we obtain $\left(
 \begin{matrix}
  x'\\
  y' \\
 \end{matrix}
 \right)=\left(
 \begin{matrix}
  0 & 1 \\
  -1 & 0  \\
 \end{matrix}
 \right)\left(
 \begin{matrix}
  x\\
  y \\
 \end{matrix}
 \right)$ and transform  $\eta_{xxx}^{\text{AFM-y}}$ to $-\eta_{y'y'y'}^{\text{AFM-x}}$ and $\sigma_{yxy}^{\text{AFM-y}}$ to $-\sigma_{x'y'x'}^{\text{AFM-x}}=\sigma_{x'x'y'}^{\text{AFM-x}}$. However, the system has no $C_{4z}$ rotation symmetry but tiny magnetic anisotropy in $x$ and $y$ direction. This magnetic anisotropy might be identified in the difference between $\eta_{LI}$ and $\sigma_{CS}$ in Fig.\ref{fig2}(c) if the numerical error in calculations can be reduced enough, e.g., the generated Wannier functions can preserve symmetry exactly.

 \begin{table*}[]
\begin{tabular}{|c|cccc|}
\hline
                                     & \multicolumn{4}{c|}{{ Symmetry operation}}                                                                                                                   \\ \cline{2-5} 
                                     & \multicolumn{2}{c|}{AFM-x}                                                                        & \multicolumn{2}{c|}{AFM-y/z}                                                 \\ \cline{2-5} 
\multirow{-3}{*}{\textbf{Quantitiy}} & \multicolumn{1}{c|}{$C_{2x}\mathcal{T}$}        & \multicolumn{1}{c|}{$M_x$}                      & \multicolumn{1}{c|}{$C_{2x}$}                   & $M_x\mathcal{T}$           \\ \hline
$\Delta_{mn}^x(k_x,k_y)$             & \multicolumn{1}{c|}{$-\Delta_{mn}^x(-k_x,k_y)$} & \multicolumn{1}{c|}{$-\Delta_{mn}^x(-k_x,k_y)$} & \multicolumn{1}{c|}{$\Delta_{mn}^x(k_x,-k_y)$}  & $\Delta_{mn}^x(k_x,-k_y)$  \\ \hline
$\Delta_{mn}^y(k_x,k_y)$             & \multicolumn{1}{c|}{$\Delta_{mn}^y(-k_x,k_y)$}  & \multicolumn{1}{c|}{$\Delta_{mn}^y(-k_x,k_y)$}  & \multicolumn{1}{c|}{$-\Delta_{mn}^y(k_x,-k_y)$} & $-\Delta_{mn}^y(k_x,-k_y)$ \\ \hline
$g_{mn}^{xx}(k_x,k_y)$               & \multicolumn{1}{c|}{$g_{mn}^{xx}(-k_x,k_y)$}    & \multicolumn{1}{c|}{$g_{mn}^{xx}(-k_x,k_y)$}    & \multicolumn{1}{c|}{$g_{mn}^{xx}(k_x,-k_y)$}    & $g_{mn}^{xx}(k_x,-k_y)$    \\ \hline
$g_{mn}^{xy}(k_x,k_y)$               & \multicolumn{1}{c|}{$-g_{mn}^{xy}(-k_x,k_y)$}   & \multicolumn{1}{c|}{$-g_{mn}^{xy}(-k_x,k_y)$}   & \multicolumn{1}{c|}{$-g_{mn}^{xy}(k_x,-k_y)$}   & $-g_{mn}^{xy}(k_x,-k_y)$   \\ \hline
$g_{mn}^{yy}(k_x,k_y)$               & \multicolumn{1}{c|}{$g_{mn}^{yy}(-k_x,k_y)$}    & \multicolumn{1}{c|}{$g_{mn}^{yy}(-k_x,k_y)$}    & \multicolumn{1}{c|}{$g_{mn}^{yy}(k_x,-k_y)$}    & $g_{mn}^{yy}(k_x,-k_y)$    \\ \hline
\end{tabular}
\caption{\label{quantitity under symmerty operation}Transformation of group velocity difference ${\Delta}^a_{m n}(\mathbf{k})$ and quantum metric $g_{mn}^{bc}(\mathbf{k})$ under $C_{2x}$, $C_{2x}\mathcal{T}$, $M_x$ and $M_x\mathcal{T}$. }
\end{table*}

    The integrand of  $\eta_{\mathrm {LI }}$ can be decomposed into three terms as shown in (\ref{coefficients of injection current}), i.e., the \textit{k}-resolved group velocity difference $\Delta_{m n}^a(\mathbf{k})$, quantum metric 2$g_{mn}^{bc}(\mathbf{k})\equiv\{r_{mn}^b(\mathbf{k}),r_{nm}^c(\mathbf{k})\}$\cite{ahn2020low,ahn2022riemannian}, and joint density of states (JDOS) $\delta(\omega_{nm}(\mathbf{k})-\omega)$. The distribution of $\Delta_{m n}^a(\mathbf{k})$ and $g_{mn}^{bc}(\mathbf{k})$ in three magnetic structures are shown in Fig.\ref{fig3} and Fig.\ref{fig4}, respectively, which follow the constrains of symmerty. For instance, $C_{2x}$ and $M_x\mathcal{T}$ in AFM-$y/z$ confine $\Delta_{m n}^x(\mathbf{k})$ and $g_{mn}^{xx,yy}(\mathbf{k})$ to be symmetric about $k_y$ axis, while $\Delta_{m n}^y(\mathbf{k})$ and $g_{mn}^{xy}(\mathbf{k})$ are confined to be antisymmetric about $k_y$ axis. These properties require $\eta_{LI}^{yyy,xxy,yxx}$ to vanish in AFM-$y/z$ structures, which is consistent with above symmetry analysis and calculation results. More details about the transformation of quantities under symmetrical operations can be found in Table.\ref{quantitity under symmerty operation}. Furthermore, $\Delta_{m n}^a(\mathbf{k})$ in the three magnetic structures are similar to each other while $g_{mn}^{ab}(\mathbf{k})$ are quite different, verifying above conclusion that the alignment of magnetic moments has small influence on the band structure but greatly changes the geometrical properties of wave functions as reflected in the nonlinear optical responses.
   
    The distribution of these quantities in BZ can offer us further information about the BPVE responses. Noticing that there are four peaks at 0.024, 0.33, 0.37 and 0.55 eV in linear injection responses as shown in Fig.\ref{fig2}(b-c), we can figure out the corresponding interband transitions in the Brillouin zone according to the energy conservation required by $\delta(\omega_{nm}-\omega)$. As illustrated in Fig.\ref{fig2}(a), the red, blue and green arrows marked the interband contributions to peaks centering at 0.33 eV, 0.37 eV and 0.55 eV, respectively. They locate around $\Gamma$ point in $\Gamma-K$ and $\Gamma-M$ directions. Besides, interband transition at 0.024 eV is right at $\Gamma$. The photocurrent peaks at 0.024 eV, 0.37 eV and 0.55 eV coincide with the peaks in JDOS at the same photon energy position as shown in Fig.\ref{figs2}, while the photocurrent peak at 0.33 eV has no corresponding peak in JDOS. We think this peak is mostly contributed by the large values of quantum metric $g_{mn}^{bc}(\mathbf{k})$ in $\Gamma-K$ direction from the interband transition marked by the red arrows in Fig.\ref{fig2}(a).

	\section{Computational Details}
	The band structures are calculated using the Quantum ESPRESSO (QE) simulation package\cite{giannozzi2009quantum,giannozzi2017advanced} with the generalized gradient approximation of Perdew-Burke-Ernzerhof exchange-correlation potential\cite{perdew1996generalized} in PSLIBRARY\cite{dal2014pseudopotentials}. The self-consistent calculations are carried out on a 11$\times$11$\times$1 Monkhorst-Pack \emph{k}-mesh with the plane-wave function cutoff set to 150 Ry. A vacuum region of about 15 Å along the z direction is adopted to eliminate the artificial layer interactions. Due to the existence of Eu-4\emph{f} orbitals, we apply a Hubbard \emph{U} correction with parameter $\emph{U}_{4f}=\emph{U-J}=5.0 \ \mathrm{eV}$. The maximally-localized Wannier functions\cite{marzari1997maximally,souza2001maximally,marzari2012maximally} are generated using \emph{p} orbital Sn, \emph{p} orbital of As, \emph{f} orbital of Eu. With the tight-binding Hamiltonian constructed from these Wannier functions, we calculate the optical responses with modified code of the WANNIER90 package\cite{mostofi2014updated}. The calculation is performed on the 1500×1500×1 k-mesh and the convergence has been tested.

    \section{Conclusion}
    In summary, we demonstrate that the BPVE photocurrents can be classified into linear/circular shift current and linear/circular injection current by analyzing their dependences on the polarization of incident light and generating mechanism. Taking bilayer $\mathrm{EuSn_2As_2}$ in three AFM structures as an example, we analyze the BPVE responses constrained by the magnetic symmetry through group representation theory and first-principles calculation. The results show that AFM-$z$ structure has only linear injection currents while AFM-$x/y$ structures have the linear injection and circular shift currents and the currents flow in perpendicular direction when illuminated by light. We design a protocol to distinguish the magnetic structures of even-layer $\mathrm{EuSn_2As_2}$ in experiments, which can be generalised to materials with similar MPGs. Moreover, we illustrate that the geometric quantity has a crucial role in the BPVE responses, which can guide us to find the ideal materials for practical applications.

	\begin{acknowledgments}
    We acknowledge the discussion with Jiacheng Gao and Yi Jiang. This work was supported by the Ministry of Science and Technology of China (Grants No. 2018YFA0305700 and No. 2022YFA1403800), the National Natural Science Foundation of China (Grants No. 11925408, No. 11921004, and No. 12188101), the Chinese Academy of Sciences (Grant No. XDB33000000) and the Informatization Plan of the Chinese Academy of Sciences (Grant No. CAS WX2021SF-0102), and the Condensed Matter Physics Data Center, CAS. 
	\end{acknowledgments}
	\bibliography{apssamp}

\clearpage
\newpage
	\setcounter{section}{0}
	\setcounter{equation}{0}
	\renewcommand{\theequation}{S\arabic{equation}}
	\renewcommand{\bibnumfmt}[1]{[S#1]}
	\renewcommand{\citenumfont}[1]{S#1}	
	\setcounter{table}{0}
	\renewcommand{\thetable}{S\arabic{table}}%
	\setcounter{figure}{0}
	\renewcommand{\thefigure}{S\arabic{figure}}%
	\appendix

	\begin{widetext}

\section{Band structures of the AFM-$x/y$ structures and JDOS.}
		\begin{figure*}[htbp]
		\centering 
		\includegraphics[width=0.8\textwidth]{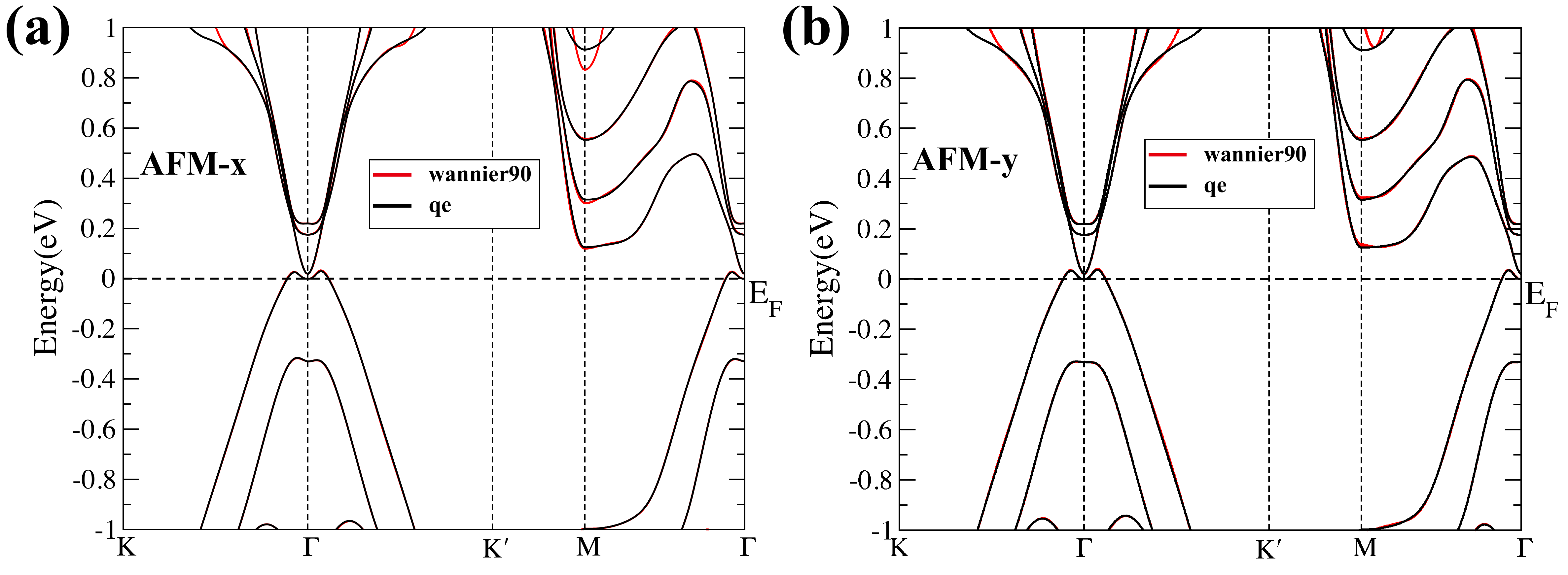}
		\caption{The band structures of bilayer $\mathrm{EuSn_2As_2}$ with spin-orbit coupling (SOC) calculated with VASP (black) and tight-binding Hamiltonian based on the Wannier functions (red) in (a) AFM-$x$ phase  and (b) AFM-$y$ phase. \label{figs1}}
	\end{figure*}
  \begin{figure*}[htbp] 
		\centering 
		\includegraphics[width=0.5\textwidth]{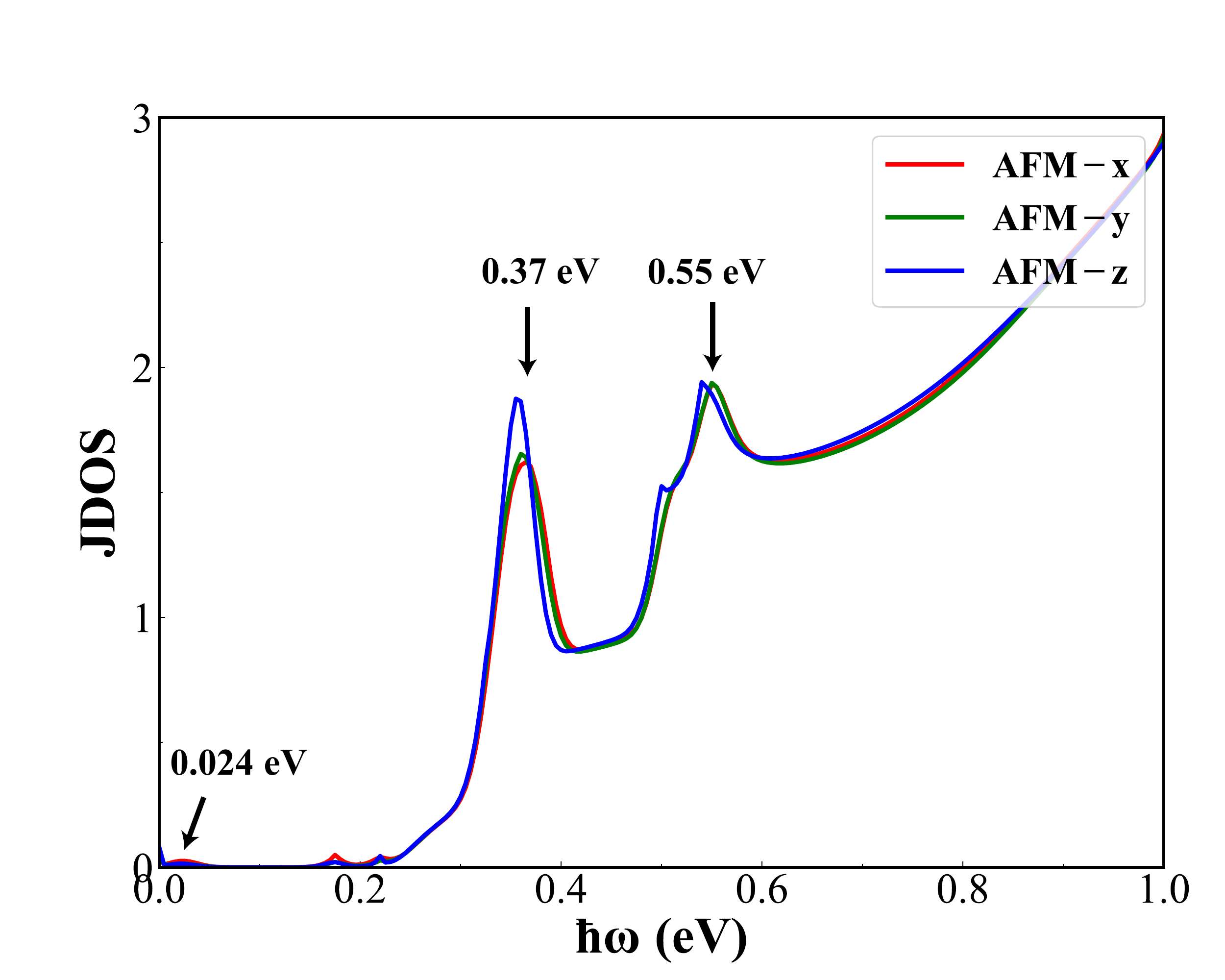}
		\caption{The joint density of states of three magnetic structures.\label{figs2}}
	\end{figure*}
\section{Symmetry analysis and character tables of the AFM-$x$ and AFM-$y$ phases}
\paragraph{\textbf{AFM-$x$ phase}}
The bilayer AFM-$x$ $\mathrm{EuSn_2As_2}$ belongs to MPG $2'/m$ that can be decomposed into $\mathcal{M}=C_s\oplus\left(C_{2h}-C_s\right)\mathcal{T}$ and $\mathcal{M}=C_s\otimes\{\mathcal{I},\mathcal{PT}\}$.	The representation of $\sigma_{\text{CS}}$ and $\eta_{\text{LI}}$ are given below according to Table. \ref{characters of Cm}. As there are one and three identity representations in $\Gamma_{\sigma_{\text{CS}}}$ and $\Gamma_{\eta_{\text{LI}}}$, respectively, we can find out the nonvanishing independent components are $\sigma_{\text{CS}}^{xxy}$ and $\eta_{\text{LI}}^{yyy},\eta_{\text{LI}}^{yxx},\eta_{\text{LI}}^{xxy}$.

\begin{equation}
	\begin{split}
		&\Gamma_{\sigma_{\text{CS}}}=\Gamma_{J}\otimes\Gamma_{(EE^{\star})_{\text{asym}}}=A_1\oplus A_2,\\
		&\Gamma_{\eta_{\text{LI}}}=\Gamma_{J}\otimes\Gamma_{(EE^{\star})_{\text{sym}}}=3A_1\oplus 3A_2.
	\end{split}
\end{equation}

\paragraph{\textbf{AFM-$y$ phase}}
The AFM-$y$ phase belongs to the MPG $2/m'$ that can be decomposed into $\mathcal{M}=C_2\oplus\left(C_{2h}-C_2\right)\mathcal{T}$ and $\mathcal{M}=C_2\otimes\{\mathcal{I},\mathcal{PT}\}$. 	The representation of $\sigma_{\text{CS}}$ and $\eta_{\text{LI}}$ are given below according to Table. \ref{characters of C2}. As there are one and three identity representations in $\Gamma_{\sigma_{\text{CS}}}$ and $\Gamma_{\eta_{\text{LI}}}$, respectively, we can find out the nonvanishing independent components are $\sigma_{\text{CS}}^{yxy}$ and $\eta_{\text{LI}}^{yxy},\eta_{\text{LI}}^{xyy},\eta_{\text{LI}}^{xxx}$.
		\begin{equation}
	\begin{split}
		&\Gamma_{\sigma_{\text{CS}}}=\Gamma_{J}\otimes\Gamma_{(EE^{\star})_{\text{asym}}}=A_1\oplus A_2,\\
		&\Gamma_{\eta_{\text{LI}}}=\Gamma_{J}\otimes\Gamma_{(EE^{\star})_{\text{sym}}}=3A_1\oplus 3A_2.
	\end{split}
\end{equation}

	\begin{table}[H]
	\centering
	\begin{tabular}{c|cc|c}
		\toprule
		$C_s$  & $E$ & $M_x$&basis function \\
		\hline
		$A_1$ & 1 & 1&$z,\ y,\ R_x$\\
		$A_2$ & 1 &-1&$x,\ R_z,\ R_y$\\
		\hline\
		$\Gamma_{(EE^{\star})_{\text{sym}}}$ & 3 & 1&\\
		$\Gamma_{(EE^{\star})_{\text{asym}}}$& 1 & -1&\\
		\bottomrule
	\end{tabular}
	\caption{\label{characters of Cm}Characters for irreducible representations of group $C_s$ and characters of a polar vector, a symmetric second-rank tensor and an asymmetric second-rank tensor. }
\end{table}

	\begin{table}[H]
	\centering
	\begin{tabular}{c|cc|c}
		\toprule
		$C_2$  & $E$ & $C_{2x}$&basis function \\
		\hline
		$A_1$ & 1 & 1&$x,\ R_x$\\
		$A_2$ & 1 &-1&$y,\ z,\ R_y,\ R_z$\\
		\hline
		$\Gamma_{(EE^{\star})_{\text{sym}}}$ & 3 & 1&\\
		$\Gamma_{(EE^{\star})_{\text{asym}}}$& 1 & -1&\\
		\bottomrule
	\end{tabular}
	\caption{\label{characters of C2}Characters for irreducible representations of group $C_2$ and characters of a polar vector, a symmetric second-rank tensor and an asymmetric second-rank tensor. }
\end{table}

\end{widetext}

\end{document}